%% file: PNAS-template-main.tex
\documentclass[9pt,twocolumn,twoside]{pnas-new}
\PassOptionsToPackage{numbers, compress}{natbib}

\usepackage{amsmath}
\usepackage{amsthm}
\usepackage{subcaption}
\usepackage{balance}
\newcommand{\orcid}[1]{\href{https://orcid.org/#1}{\textcolor[HTML]{A6CE39}{\aiOrcid}}}

\templatetype{pnasresearcharticle} % Choose template 
\title{Analyzing the Design Space of Re-opening Policies and COVID-19 Outcomes in the US}

\author[a,1]{Chaoqi Yang}
\author[a,1]{Ruijie Wang} 
\author[a]{Fangwei Gao}
\author[a]{Dachun Sun}
\author[a]{Jiawei Tang} 
\author[a,2]{Tarek Abdelzaher}

\affil[a]{University of Illinois, Urbana-Champaign, IL 61801, USA}

\leadauthor{Yang} 

\significancestatement{The work predicts the impact  of a wide range of social distancing and reopening policies on COVID-19 propagation outcomes in different communities. Candidate policies manipulate the availability of specific social mixing domains, such as closure/resumption of business for some part of the workforce, changes in business opening hours and/or in maximum allowable occupancy, closure/resumption of schools, cancellation of events above a specified size, and/or imposing testing requirements as a precondition of entry into various establishments. The produced estimates allow making informed policy choices in the face of pressure to relax distancing restrictions at a time when no effective treatment or vaccine are widely available. Initial exploration of the space of partial re-opening policies suggests that mandatory testing may be a more effective strategy than social distancing. 
}

\authorcontributions{Author contributions: C.Y., R.W. and T.A. designed research; C.Y., R.W. and T.A. performed research; C.Y., R.W., F.G., D.S. and J.T. constructed the website; C.Y., R.W. and T.A. wrote the paper.}
\authordeclaration{The authors declare no competing interest.}
\equalauthors{\textsuperscript{1}C.Y. contributed equally to this work with R.W..}
\correspondingauthor{\textsuperscript{2}To whom correspondence should be addressed. E-mail: zaher@illinois.edu.}


\input{10-abstract}

\dates{This manuscript was compiled on \today}
\begin{document}

\maketitle
\thispagestyle{firststyle}
\ifthenelse{\boolean{shortarticle}}{\ifthenelse{\boolean{singlecolumn}}{\abscontentformatted}{\abscontent}}{}

% If your first paragraph (i.e. with the \dropcap) contains a list environment (quote, quotation, theorem, definition, enumerate, itemize...), the line after the list may have some extra indentation. If this is the case, add \parshape=0 to the end of the list environment.

\input{20-introduction}
\input{35-result}

%
%\input{background}
%

\input{50-experiment}

%
%\input{oversmooth}
%
%\input{experiment}
%
\input{30-related}

\input{60-conclusion}

% \input{70-extraresults}

% Bibliography
\bibstyle{pnas-bib}
\bibliography{PNAS-template-main}
\input{40-model}

\end{document}

%% file: 10-abstract.tex
\begin{abstract}
Recent re-opening policies in the US, following a period of social distancing measures, introduced a significant increase in daily COVID-19 infections, calling for a roll-back or substantial revisiting of these policies in many states. The situation is suggestive of difficulties modeling the impact of partial distancing/re-opening policies on future epidemic spread for purposes of choosing safe alternatives. More specifically, one needs to understand the impact of manipulating the availability of social interaction venues (e.g., schools, workplaces, and retail establishments) on virus spread. We introduce a model, inspired by social networks research, that answers the above question. Our model compartmentalizes interaction venues into categories we call {\em mixing domains\/}, enabling one to 
predict COVID-19 contagion trends in different geographic regions under different ``what if'' assumptions on partial re-opening of individual domains. %Predictions show that if social distancing was removed at the time of this writing, XXX individuals will acquire the disease by YYY. A partial return of only 50\% of the currently affected workforce will reduce that number to ZZZ. Without a treatment or vaccine, given current conditions, the virus will infect ZZZ individuals by ZZZ. 
We apply our model to several highly impacted states showing (i) how accurately it predicts the extent of current resurgence (from available policy descriptions), and (ii) what alternatives might be more effective at mitigating the second wave. We further compare policies that rely on partial venue closure to policies that espouse wide-spread periodic testing instead (i.e., in lieu of social distancing). Our models predict that the benefits of (mandatory) testing out-shadow the benefits of partial venue closure, suggesting that perhaps more efforts should be directed to such a mitigation strategy.
\end{abstract}

%% file: 20-introduction.tex
\section{Introduction}
\dropcap{A}ccording to the International Monetary Fund (IMF), the COVID-19 pandemic could cost the world economy up to \$9 Trillion USD, nearly the combined gross domestic product (GDP) of Japan and Germany, or roughly half that of the US.\footnote{https://blogs.imf.org/2020/04/14/} Since the first official case was confirmed in Wuhan, China, more than 16 million individuals have been infected at the time of writing, leading to over 650 thousand deaths.\footnote{https://coronavirus.jhu.edu} Compared to such diseases as the 2003 SARS-CoV, the H1N1 influenza A, and the Ebola virus, COVID-19 shows strong infectiousness, with a reproductive number, $R_0 > 3$, according to some studies~\cite{liu2020reproductive}. In the absence of a widely-available treatment or vaccines, social distancing is thought to be an effective strategy to protect from the virus~\cite{NPIS,Intervention}. Prolonged social distancing has other side-effects, both socially and economically~\cite{BROOKS2020912,NBERw26867}, leading to pressure to relax stricter policies. It is important to reach a finer-grained understanding of available crisis management options that minimize side-effects of re-opening.

We develop algorithms that model the impact of different {\em venue-centric partial re-opening policies\/} on the future spread of COVID-19. By {\em venue-centric\/}, we refer to policies that manipulate availability, occupancy, opening hours, or admission conditions in a venue-dependent manner. This granularity of policy definition allows expression of finer-grained policies, such as ones that manipulate occupancy of specific types of venues; for example, re-open the retail sector but not schools, or limit restaurant occupancy to a given fraction of capacity. Given data collected under some policies, the algorithm predicts the impact of other potential policies to implement next. The goal is to answer the question of ``return to normal''. What steps are safe to take towards restoration of elements of normalcy? How to customize the answers to the special circumstances of different regional populations?

The work borrows insights from information cascade propagation on social networks~\cite{abdelzaher2020multiscale}. Information (similarly to viral contagion) propagates through {\em broadcast channels\/}. In the information space, a broadcast channel might be a Facebook wall, online subreddit, or virtual “hangout”. These virtual spaces create opportunities for information transmission among individuals who frequent them. 
In the world of viral contagion, physical spaces, such as stores, offices, public transport, and family residences, constitute social interaction venues that serve the role of broadcast channels. It is useful to think of {\em categories\/} of such spaces, such as schools, workplaces, and brick-and-mortar retail shops. We call these categories, social {\em mixing domains\/}. Venue-centric social distancing policies manipulate the availability, hours, and entry conditions of some of these domains. Information cascade models predict what happens to propagation when the underlying broadcast channels are manipulated. Leveraging this analogy, we capture the effects of a wide playbook of social distancing and reopening options on contagion dynamics. 

%% file: 35-result.tex
\section{Venue-based Propagation Decomposition}
\label{sec:result}
The main contribution of this work is a model and its application (to COVID-19) that decomposes the equivalent overall virus transmissibility as a weighted sum over social interaction venues in order to understand the contribution of different interaction venues to the overall transmission dynamics and enable exploration of a space of partial closure/re-opening policies that manipulate availability of social interaction venues to minimize spread. The results stem from the {\em propagation decomposition theorem\/}, derived by the authors and stated below, allowing one to investigate the impact of venue-based COVID-19 mitigation strategies. The proof of the theorem is shown in Appendix~A.

\begin{adjustwidth}{1cm}{}

\medskip
{\bf The Propagation Decomposition Theorem:\/} Consider a geographic region where the total population comprises, for analysis purposes, a set $\mathcal{G}$ of non-overlapping groups (e.g., employed and unemployed, or minors, adults, and seniors).  Individuals split their time among $n$ social {\em interaction venues\/}, such that individuals in group $g_j \in \mathcal{G}$ spend, on average, a fraction, $\eta_{ij}$ of their time in venue $i$ (some fractions could be zero). Let $\alpha_i$ be the nominal occupancy of venue $i$, normalized to population size. We shall henceforth call it, venue {\em size\/}. Let $f_{ij}$ be the average fraction of occupancy of venue $i$ who are from group $g_j$, and Let $\tau_i$ be the probability that an infected person in venue $i$ infects another in the same venue (i.e., venue-specific transmissibility). Then, the equivalent overall transmissibility of the virus in the community is: 

\begin{equation}
    \tau_{eq} =  \sum_{i=1}^n \sum_{j \in \mathcal{G}} \tau_i ~f_{ij} ~  \eta_{ij}   ~\alpha_i^2
    \label{eq:theorem}    
\end{equation}

\end{adjustwidth}

\noindent
The above result is derived by analyzing the interactions between the susceptible and the infected individuals. It is valid for epidemiological models that include the {\em susceptible\/} and {\em infected\/} states, such as the SIS, SIR, and SEIR models, as well as their extensions. Its proof is described in Appendix~A. Table~\ref{tab:terminology} defines the used terminology.
%We also demonstrate the application of this result, as well as several simplified forms thereof (corollaries) that offer easier-to-compute approximate answers. 
%In the following, we illustrate the result in the context of an SIR model. The same theorem holds in other models, such as SIS, and SEIR. In the context of an SIR model, the predicted $\tau_{eq}$, computed by the the mixing theorem is used in the following equations:

%\begin{align}
%    \frac{d S(t)}{dt} &= - \tau_{eq}\cdot S(t)I(t), \nonumber \\
%    \frac{d I(t)}{dt} &= \tau_{eq}\cdot S(t)I(t) - \gamma \cdot I(t), \nonumber\\
%    \frac{d R(t)}{dt} &= \gamma \cdot I(t), \nonumber
%\end{align}

%\noindent
%where $S(t)$, $I(t)$, $R(t)$, $\gamma$ and $N$ are the susceptible, infected, recovered/removed, the recovery rate and the total population, respectively. 

\begin{table}[ht]
    \centering
    \caption{Terminology Used}    \begin{tabular}{c|l}
    \toprule
        $N$ & Total community size. \\ \midrule
        $n$ & Total number of mixing domains. \\ \midrule
        $\gamma$ & Recovery rate \\ \midrule
        $S(t)$ & Total number of susceptible individuals at time $t$. \\ \midrule
        $I(t)$ & Total number of infected individuals at time $t$. \\ \midrule   $R(t)$ & Total number of recovered individuals at time $t$. \\ \midrule   
        $D_i$ & The $i$-th Mixing domain.\\ \midrule
        $N_i$ & Number of members in domain $D_i$ \\ \midrule
        $\beta_i$ & Per person rate of spread (over all encounters) in \\ &domain $D_i$,
        per unit time \\ \midrule
        $\tau_i$ & Transmissibility per encounter in domain $D_i$, per \\ &unit time \\ \midrule
        $\eta_i$ & The average time a member spends in domain $D_i$. \\ \midrule
        $S_i(t)$ & Expected number of susceptible individuals in domain \\&$D_i$ at time $t$.\\ \midrule
        $I_i(t)$ & Expected number of infected individuals in domain $D_i$ \\&at time $t$.\\ \midrule
        $\zeta_i(t)$ & The ratio $S_i(t)/S(t)$. \\ \midrule
        $\kappa_i(t)$ & The ratio $I_i(t)/I(t)$. \\ \midrule
        $\alpha_i$ & The ratio $N_i/N$. \\ \bottomrule
    \end{tabular}

    \label{tab:terminology}
\end{table}

%\medskip{}

%{\bf Significance:\/} {\em The result allows predicting the impact (on COVID-19 propagation in different communities) resulting from specified changes in social distancing policies that manipulate the availability of specific mixing domains, such as closure/resumption of business for some part of the workforce, changes in business opening hours, closure/resumption of schools, or cancellation of events above a specified size. The produced estimates allow making more informed policy choices as pressure mounts to relax some of the current distancing restrictions, while no treatment or vaccine are available. \/}

%% file: 50-experiment.tex
\section{Applying the Venue-based Decomposition Model}
\label{sec:evaluation}
%The proposed model is designed to foresee/assess the potential impact of new policies on transmissibility of COVID-19. The contribution is to learn a prediction model from previous observation periods whose predictions apply to a future policy change. 
%We discuss how to apply Equation~(\ref{eq:theorem}) for purposes of such prediction to regions of the US.
%
To apply Equation~(\ref{eq:theorem}), we divide the population into $G$ different groups and divide all social interaction venues into $C$ different categories, we call {\em mixing domains\/}. 
%We should also determine (or anticipate) the fraction of time each group spends in each domain. 
The model does not dictate how to break up the population and interaction venues into categories. This freedom can be exploited to construct progressively more nuanced policies that include a larger number of groups and domains. For now, we group social interaction venues into $C=4$ categories (or mixing domains), namely: {\em home\/}, {\em work\/}, {\em school\/}, and {\em commercial\/}. In general, the mixing domains should match those that a policy might wish to individually manipulate. For example, if the policy will make explicit decisions on opening/closure of venues of worship, then {\em worship\/} should be cast as an explicit mixing domain. Similarly, we categorize the population into $G=4$ age groups: {\em pre-school\/}, {\em child/teen\/}, {\em adult\/}, and  {\em senior\/}. Equation~(\ref{eq:theorem}) can now be re-written:
\begin{align}
    \tau_{eq}
    &= \sum_{k=1}^C\sum_{i=1}^{n_k}\sum_{j=1}^G\tau_i f_{kj}\eta_{kj}\alpha_i^2 \notag\\
    &= \sum_{k=1}^C\sum_{j=1}^G f_{kj} \eta_{kj}\sum_{i=1}^{n_k}\tau_i\alpha_i^2 
    \label{eq:final}
\end{align}
In Equation~(\ref{eq:final}), $k$ enumerates the four possible mixing domains: {\em home\/}, {\em work\/}, {\em school\/}, and {\em commercial\/}; $i$ enumerates the specific social interaction venues included in each mixing domain; and $j$ is the age group indicator: {\em pre-school\/}, {\em child/teen\/}, {\em adult\/}, and {\em senior\/}. Additionally, $\tau_i$ denotes transmissiblity within the $i$-th specific venue; $\alpha_i$ denotes the average occupancy of the $i$-th venue; $f_{kj}$ denotes the fraction of individuals in domain $k$ who are of group $j$; and $\eta_{kj}$ denotes the fraction of time that members of group $j$ spend in domain $k$.

\subsection{The Transmissibility Approximation}
We assume that
\textit{in a particular venue, the transmissibility is proportional to the venue size raised to power $-M$}. In other words: 

\begin{equation}
\tau_i \approx \kappa \alpha_i^{-M}
\label{eq:tau-M}
\end{equation}

\noindent
where $M>0$ and $\kappa$ is a proportionality constant. Smaller venues (e.g., a family residence) typically feature closer interactions than larger ones. Thus, while in a larger venue, the expected number of individuals that a person might infect is {\em larger\/}, the single person-to-person transmission probability may be higher in a smaller venue. Accordingly, substituting from  Equation~(\ref{eq:tau-M}) into Equation~(\ref{eq:final}), we get the following approximation:

\begin{equation}
    \tau_{eq} \approx \kappa \sum_{k=1}^C\sum_{j=1}^G f_{kj} \eta_{kj}\sum_{i=1}^{n_k}\alpha_i^{2-M} 
    \label{eq:approx-tarek}
\end{equation}

\noindent
The parameter, $M$, and the proportionality constant, $\kappa$, can be empirically estimated from observational data as shown in Section~\ref{sec:outcome}. Next, we describe another key approximation.

\subsection{The Zipf Approximation}
A challenge in Equation~(\ref{eq:approx-tarek}) is to carry out the last summation, since it ranges over all individual interaction venues in mixing domain, $k$.
In practice, it is not always possible (or at least very cumbersome) to determine the size of all the individual venues in a given mixing domain (e.g., all residences within the {\em home\/} domain). However, we usually know the mean, $Mean_k$. Thus, we can use a distribution. Let the venue sizes in domain $k$ be given by the distribution, $p_k (x)$, over the size variable, $x$. In this paper, we assume that venue sizes are given by a power-law (i.e., Zipf) distribution. This distribution is commonly found in nature to describe popularity, a concept that is quite related to occupancy. Thus, consider the distribution $p_k (x) = a_k x^{-b_k}$, where $lower_k \leq x \leq upper_k$, and $p_k(x)=0$ elsewhere. We then have:

\begin{align}
    \int_{lower_k}^{upper_k} p_k(x)\cdot dx &= 1 \notag\\
    \int_{lower_k}^{upper_k} x\cdot p_k(x) \cdot dx &= Mean_k. 
    \label{eq:avg}
\end{align}

\noindent
The limits $upper_k$ and $lower_k$ are the largest and smallest venue size (referring to nominal occupancy of the venue) within domain $k$. Substituting with $p_k (x) = a_k x^{-b_k}$ in the above two equations, $a_k$ and $b_k$ can be found, thus completely specifying the distribution of venue sizes in the domain. Equation~(\ref{eq:approx-tarek}) then becomes:

\begin{equation}
    \tau_{eq} \approx \kappa \sum_{k=1}^C\sum_{j=1}^G f_{kj} \eta_{kj} \int_{lower_k}^{upper_k} a_k x^{-b_k} \cdot x^{2-M} dx
    \label{eq:approx2-tarek}
\end{equation}

\subsection{Policy Specification} 
From Equation~(\ref{eq:approx2-tarek}),
to compute the impact of a new policy,  $\mathcal{P}$, on overall virus transmissibility, one needs to estimate the values of the product $f_{kj} \eta_{kj}$ (for each social group $j$ and mixing domain $k$) under the new policy, as well as the distribution $p_k(x)$ (given by parameters $a_k$, $b_k$, $lower_k$ and $upper_k$, described above). Let us denote these policy-specific parameters by the tuple 
$v^{(\mathcal{P})} = \langle f_{kj}^{(\mathcal{P})}\eta_{kj}^{(\mathcal{P})}, p_k(x)^{(\mathcal{P})} \rangle$. Note that, the product $f_{kj}^{(\mathcal{P})}\eta_{kj}^{(\mathcal{P})}$ is the fraction of person-hours spent in domain $k$ by members of group $j$. Substituting with these parameters in Equation~(\ref{eq:approx2-tarek}), one gets the corresponding transmissibility $\tau_{eq}^{(\mathcal{P})}$. We are now ready for outcome prediction, described below.

\subsection{Outcome Prediction}
\label{sec:outcome}
To compute $M$ and $\kappa$ one needs to fit past known infection incidence data (in order to empirically measure transmissibility) under two different previously implemented policies, say $\mathcal{P}_1$ and $\mathcal{P}_2$, producing two instances of Equation~(\ref{eq:approx2-tarek}), one for each past policy. Since the parameters, $v^{(\mathcal{P}_1)}$ and $v^{(\mathcal{P}_2)}$ are known for previously implemented policies, the only unknowns are $\kappa$ and $M$ that can then be determined. We call these two previously implemented policies, {\em calibration policies\/}. To predict outcomes of a new policy, $\mathcal{P}_3$, one can now substitute the previously computed, $M$ and $\kappa$, together with the policy description, $v^{(\mathcal{P}_3)}$, in Equation~(\ref{eq:approx2-tarek}) to compute the corresponding new transmissibility. In most regions in the US, the above approach is feasible because several policies have already been attempted. For example, most regions were initially under a no-distancing policy (i.e., $\mathcal{P}_1 =$ absence of a distancing policy), then they were under a strict distancing policy, say, $\mathcal{P}_2$. The question they are grappling with is the re-opening policy, say, $\mathcal{P}_3$. Thus,
prediction is done as follows:
\begin{itemize}
    \item {\bf Step 1:\/} Fit an epidemiological model to observational infection incidence data observed under (at least) two different policies, $\mathcal{P}_1$ and $\mathcal{P}_2$, in order to empirically determine $\tau_{eq}^{(\mathcal{P}_1)}$ for the first policy and $\tau_{eq}^{(\mathcal{P}_2)}$ for the second policy (from curve-fitting). Let the empirically found ratio $\tau_{eq}^{(\mathcal{P}_1)}/\tau_{eq}^{(\mathcal{P}_2)}$ be denoted by $c$.
    \item {\bf Step 2:\/} Use demographic data to compute the parameter sets: $v^{(\mathcal{P}_1)}$ and $v^{(\mathcal{P}_2)}$ for these previous policies.
    \item {\bf Step 3:\/} Estimate $M$ from the empirically found ratio of values of transmissibility: $\tau_{eq}^{(\mathcal{P}_1)}/\tau_{eq}^{(\mathcal{P}_2)} = c$. Specifically, substitute on the right-hand-size from Equation~(\ref{eq:approx2-tarek}) (with the corresponding values of $v^{(\mathcal{P}_1)}$ and $v^{(\mathcal{P}_2)}$) to replace $\tau^{(\mathcal{P}_1)}_{eq}$ and $\tau^{(\mathcal{P}_2)}_{eq}$, respectively.  The only unknown in the resulting equation should be $M$, which can thus be determined. As mentioned above, one can think of policies $\mathcal{P}_1$ and $\mathcal{P}_2$ as {\em calibration\/} policies.
    \item {\bf Step 4:\/} For a new policy, $\mathcal{P}_3$, whose effect is to be predicted, calculate the transmissibility $\tau_{eq}^{(\mathcal{P}_3)}$ using the estimated $M$ and parameters $v^{(\mathcal{P}_3)}$. As before, to avoid $\kappa$, compute a ratio such as $\tau_{eq}^{(\mathcal{P}_3)}/\tau_{eq}^{(\mathcal{P}_2)}$ from Equation~(\ref{eq:approx2-tarek}), then use the empirically found $\tau_{eq}^{(\mathcal{P}_2)}$ (and the computed ratio) to determine  $\tau_{eq}^{(\mathcal{P}_3)}$. 
    \item {\bf Step 5:\/}
    The value of $\tau_{eq}^{(\mathcal{P}_3)}$ can then be used in an epidemiological model to predict the future infection incidence time series. In the rest of this work, we shall use a modified SIR that features a decaying function $f(t)$ to emulate gradual transition among policies (see Appendix~B
    for details) to ensure the smoothness:

\begin{align}
    \frac{d S(t)}{dt} &= -\tau_{eq}^{(\mathcal{P})}\cdot f(t)\cdot S(t)I(t), \\
    \frac{d I(t)}{dt} &= \tau_{eq}^{(\mathcal{P})} \cdot f(t)\cdot S(t)I(t) - \gamma \cdot I(t),\\
    \frac{d R(t)}{dt} &= \gamma \cdot I(t), 
\end{align}

\noindent
where $S(t)$, $I(t)$, and $R(t)$ denote the numbers of susceptible, infected, and recovered individuals at time $t$, whereas $\gamma$ denotes the recovery rate. 
\end{itemize}

\section{An Illustrative Example of COVID-19 Predictions}
Next, we offer an example of quantifying $\tau_{eq}^{(\mathcal{P})}$ for a specific state in the US with policy, $\mathcal{P}$. For the initial calibration policies, we use the ``no-distancing'' policy (i.e., the initial situation before any distancing policies went into effect) and the first distancing policy that was implemented (which we call ``strict distancing''). To illustrate, we pick California (CA) and Georgia (GA). Other states follow a similar process.

\subsection{Demographic and Domain Data Collection}
To populate the vector, $v^{(\mathcal{P})}$, for policy, $\mathcal{P}$, the following data is required (per state): (i) population distribution by age group; (ii) daily distribution of time across mixing domains, per age group; (iii) distribution of venue sizes within each mixing domain category.

\subsubsection{(i) Population Distribution} 
We divide the population into four age groups in this example: pre-school (<6), child/teen (6-19), adult (20-64), and senior (>64), {\em roughly\/} correlated with time distribution across different domains (pre-school kids, school age individuals, working adults, and retirees). The raw data are collected from US Census Bureau \footnote{https://www.census.gov/quickfacts/fact/table/US/PST045219}. We report the percentage in Table~\ref{tb:population_structure}.

\begin{table}[htbp]
\centering
\caption{Population Structure in California (CA) and Georgia (GA)}
\begin{tabular}{c|cccc}
\toprule
& Pre-school (<6) & Teen (6-19) & {Adult (20-64)} & Senior (>64) \\
\midrule
CA & 7.1\% & 15.3\% & 62.0  \%   & 15.6  \%  \\
GA & 7.5\% & 16.3\%  & 62.3 \%     & 13.9  \%  \\ \bottomrule
\end{tabular}
\label{tb:population_structure}
\end{table}

\subsubsection{(ii) Time Distribution} 
For each age group, the daily time structure could be found in US Bureau of Labor Statistics\footnote{https://www.bls.gov/news.release/pdf/atus.pdf}. In this data source, the time structure is allocated by ``primary activities''. In our experiment, we group these ``primary activities'' by their locations, roughly corresponding to our four major categories: home, work, school, and commercial. The statistics 
before social distancing (i.e., for $\mathcal{P}$ = ``no distancing'') are shown in Table~\ref{tb:age_group}.

\begin{table}[htbp]
\centering
\caption{Time Structure for Age Groups (Before Social Distancing)}
\begin{tabular}{c|cccc}
\toprule
 unit (h) & {Home} & {Work} & {School} & {Commercial} \\
\midrule
pre-school                 & 24.00                 & 0.00                    & 0.00                       & 0.00                        \\
teen              & 17.83                    & 1.41                     & 3.06                       & 1.70                          \\
adult                & 17.01                    & 3.68                     & 1.70                        & 1.61                           \\
senior                  & 21.42                   & {0.77} & {0.02}  & 1.80  \\           \bottomrule
\end{tabular}
\label{tb:age_group}
\end{table}

\subsubsection{(iii) Initial Social Distancing Policy in CA and GA}
The first social distancing policy ($\mathcal{P}$ = ``strict distancing'') changed the population's time distribution profile. Schools were fully closed. Non-essential workers were ordered to stay home. This reduced the total number of individuals who report to work by 80\%  for Georgia, and 50\% for California. The commercial sector activity decreased by 50\% and 80\% (according to Google Mobility reports\footnote{https://www.google.com/covid19/mobility}) for CA and GA, respectively. Therefore, using these percentages, the new average time structure for each age group is shown in Table~\ref{tb:age_group_after}.

\begin{table}[htbp]
\centering
\caption{Time Structure (After Social Distancing) for CA/GA}
\label{tb:age_group_after}
\begin{tabular}{c|cccc}
\toprule
unit (h) & {Home} & {Work} & {School} & {Commercial} \\
\midrule
pre-school   & 24.00 / 24.00  & 0.00 / 0.00    & 0.00 / 0.00      & 0.00 / 0.00    \\
teen    & 22.44 / 22.95  & 0.71 / 0.71    & 0.00 / 0.00     & 0.85 / 0.34     \\
adult   & 21.35 / 21.84 & 1.84 / 1.84    & 0.00 / 0.00     & 0.81 / 0.32     \\
senior     & 22.72 / 23.25  & 0.39 / 0.39    & 0.00 / 0.00       & 0.90 / 0.36     \\           
\bottomrule
\end{tabular}
\end{table}

\subsubsection{(iv) Distribution of Mixing Domains} 
\label{sec:size}
As stated above, interaction venues have been divided into four main categories. Within each category, we assume that venue sizes follow the Zipf's distribution. This distribution is widely applied to popularity-related statistics in physical and social sciences, such as word usage counts in natural language~\cite{piantadosi2014zipf}, and sizes distributions of commercial firms~\cite{axtell2001zipf}. The values of $upper_k$, $lower_k$, and $Mean_k$ used for different mixing domains are reported in Table~\ref{tb:broadcast_size}.
This completes the necessary inputs that allow us to follow the algorithm described earlier in 
this section.

\begin{table}[htbp]
\centering
\caption{Upper/Lower/Mean value for Mixing Domain Distribution}
\begin{tabular}{c|cccc}
\toprule
& Home \footnote{https://www.census.gov/quickfacts} & Work \footnote{https://www.quora.com/Coworking-What-is-good-space-size-people-number-ratio} & School \footnote{https://nces.ed.gov/surveys/sass/tables/sass0708\_2009324\_t1s\_08.asp} & Commercial \\
\midrule
Mean (CA) & 2.49 & 100  & 26.2   & 42         \\
Mean (GA)   & 2.33 & 100  & 28.2   & 42      \\
Upper & 7 & 1,000 & 200 & 1,000\\
Lower & 2 & 2 & 10 & 10 \\
\bottomrule
\end{tabular}\label{tb:broadcast_size}
\end{table}

\subsection{Prediction}
We continue our illustrative example, consider the following re-opening policy action categories whose impact we might want to predict:

\begin{itemize}
    \item \textbf{Policy A}: Reopen $50\%$ classes in schools. 
    \item \textbf{Policy B}: Reopen small classes in schools with maximum allowable capacity as $50$. 
    \item \textbf{Policy C}: Reopen $75\%$ offices in workplaces.
    \item \textbf{Policy D}: Reopen small offices in workplaces with maximum allowable capacity as $50$.
    \item \textbf{Policy E}: Reopen all commercial retail with maximum $75\%$ opening time.
    \item \textbf{Policy F}: Reopen all commercial retail with maximum allowable occupancy as $50$.
\end{itemize}

\noindent
Table~\ref{tb:whatif} shows the predictions for each virtual policy. While these results are actual values computed using our model, the purpose of this section remains to offer an illustrative example. Model validation is presented in the next section. A quantitative way to evaluate the safety of each policy is to compute the basic reproductive number $R_0$ of the virus under the given policy. Note that:

\begin{equation}
R_0 = \frac {\beta} {\gamma} = \frac{\tau_{eq} \cdot N}{\gamma}
\label{eq:ro}
\end{equation}

\noindent
where $\beta$ is the effective contact rate with infected individuals, $N$ is the total population, and $\gamma$ is the recovery rate. Generally speaking, policies with $R_0 < 1$ are safe, while those with $R_0 > 1$ are likely to stimulate large contagions. 

Among the example policies considered in our illustration, we find that the situation for California is worse. The values of $R_0$ for all considered policies are larger than $1$. As for Georgia, apart from reopening schools, other policies we consider in our example are safe. Policies that restrict the maximum allowable capacity of each domain seem to do better. Reopening schools, seems to have the most severe consequence.

\begin{table}[h]
\caption{Prediction after one month under different reopening policy.}
\label{tb:whatif}
\centering
\small
\begin{tabular}{l|c|c|c|c}
\hline
States  & \multicolumn{2}{c|}{California (by 7/31)} & \multicolumn{2}{c}{Georgia (by 7/31)} \\ \hline
Policy   & $R_0$       & Predictions     & $R_0$     & Predictions   \\ \hline
Strict Distancing & 1.03 & 433k & 0.72 & 165k \\
%Current Policy & 1.12 & 462k & 0.97 & 109k \\
No distancing & 1.83 & 1066k & 2.37 & 954k \\
A (50\% Schools) & 1.28      & 528k               & 1.23    & 211k              \\
B (Small Classes) & 1.26      & 519k              & 1.01    & 186k              \\
C (75\% Offices) & 1.20     & 492k                & 0.92    & 178k              \\
D (Small Offices) & 1.13      & 465k                & 0.85    & 173k              \\
E (Retail 75\%) & 1.24    & 510k               & 0.95    & 181k              \\
F (Retail 50\%) & 1.15      & 473k               & 0.87   & 174k            \\ \hline
\end{tabular}
\end{table}

%The detailed procedures include collecting sate-specific data, carrying out predictions after the new policy (this step is agnostic to ground truth data) and validate the prediction result using afterwards real COVID-19 statistics. In this paper, we divide the whole period (from the beginning XXX to today XXX) as three stages: before social distancing (XXX - XXX), after social distancing (XXX - XXX) and partially reopening (XXX - XXX). 

%In this paper, the first two periods are used as training. We validate on the partially reopening period and also provide a variety of predictions for potential future policies (e.g., fully reopening).

\begin{figure*}[ht]
\centering
\begin{subfigure}{.45\textwidth}
  \centering
  % include second image
  \includegraphics[width=1.\linewidth]{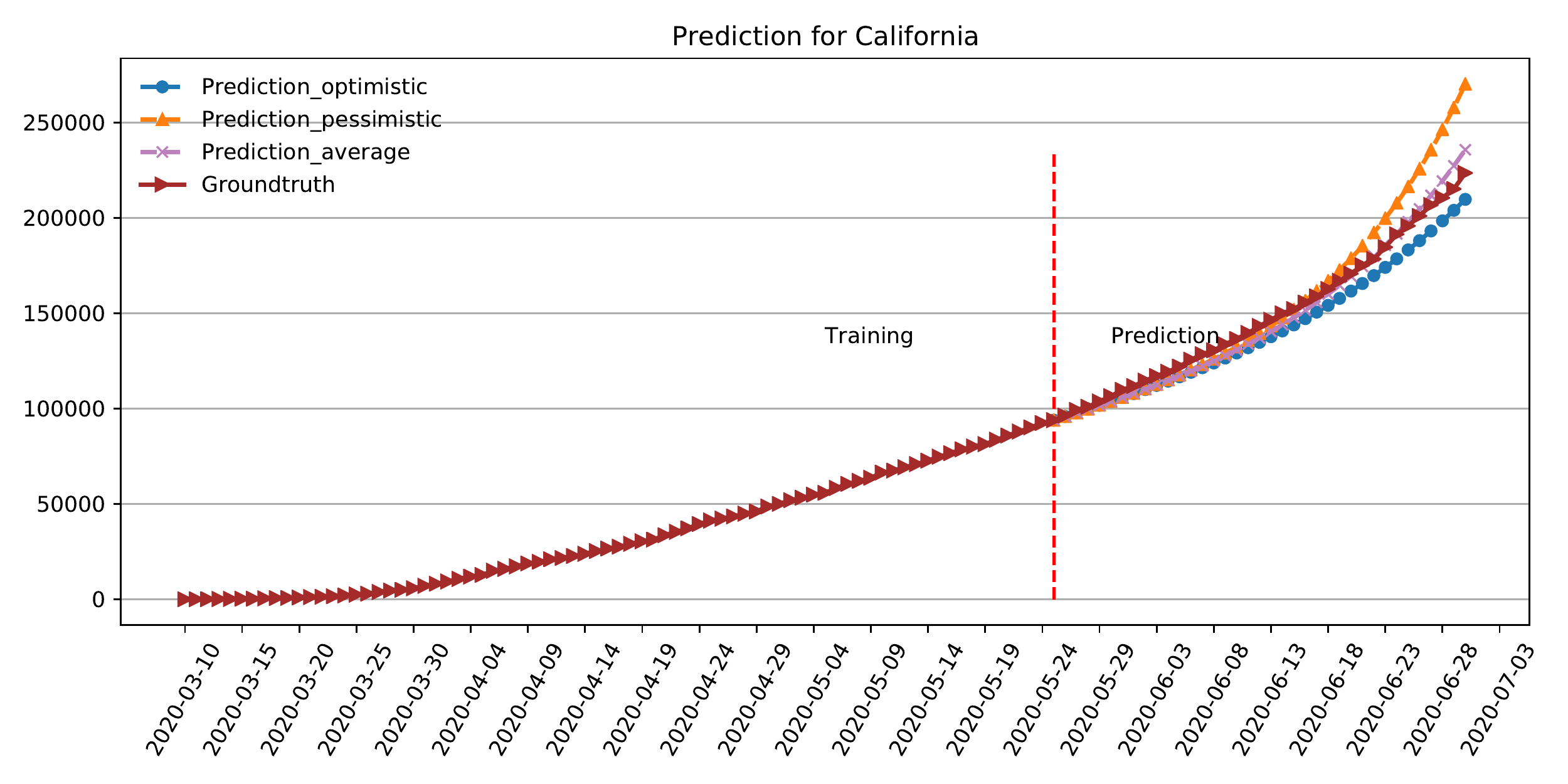}  
  \caption{California. M = $0.96$.}
  \label{fig:sub-second1}
\end{subfigure}
\begin{subfigure}{.45\textwidth}
  \centering
  % include first image
  \includegraphics[width=1.\linewidth]{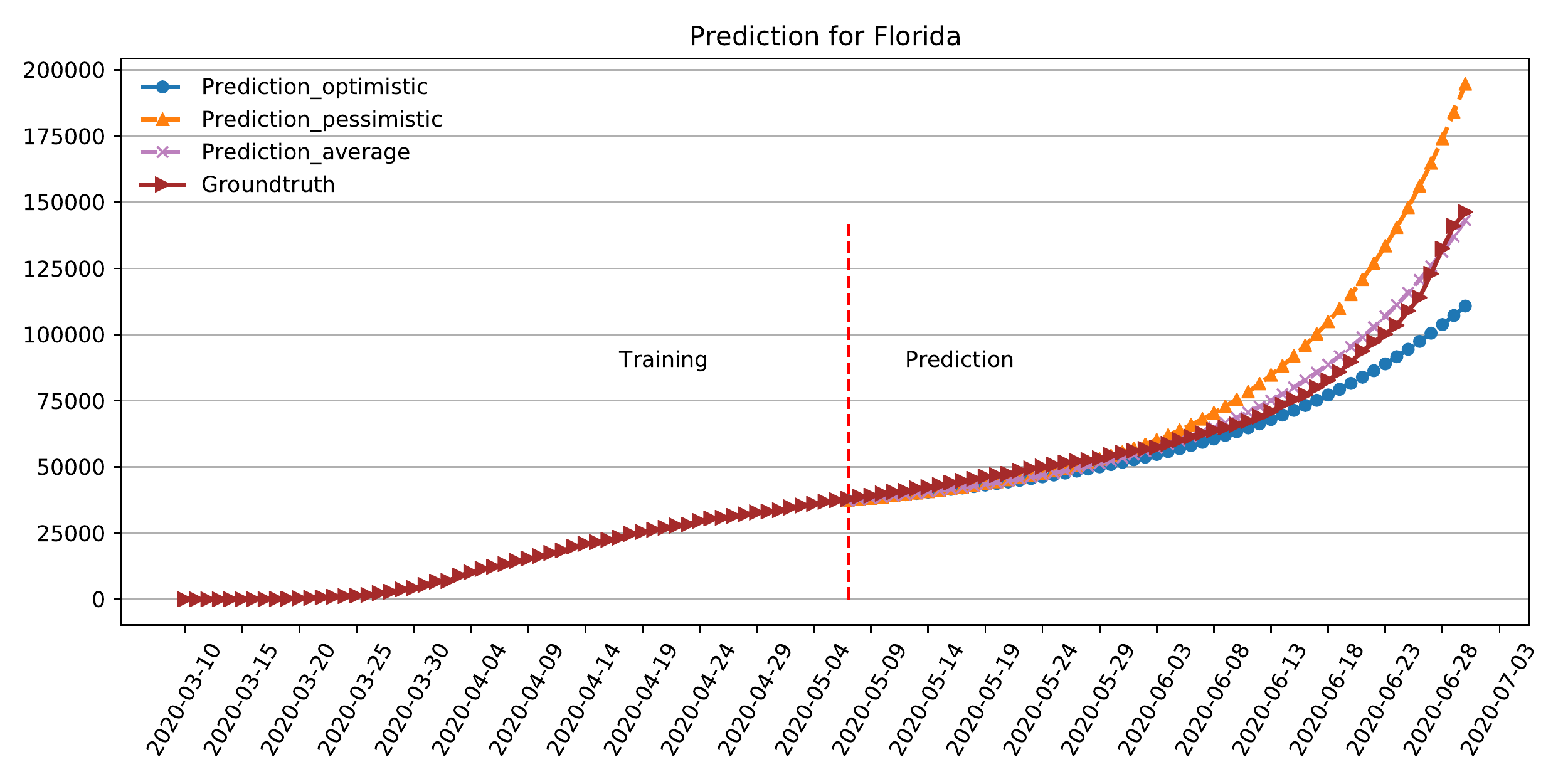}  
  \caption{Florida. M = $0.63$.}
  \label{fig:sub-first2}
\end{subfigure}

\begin{subfigure}{.45\textwidth}
  \centering
  % include second image
  \includegraphics[width=1.\linewidth]{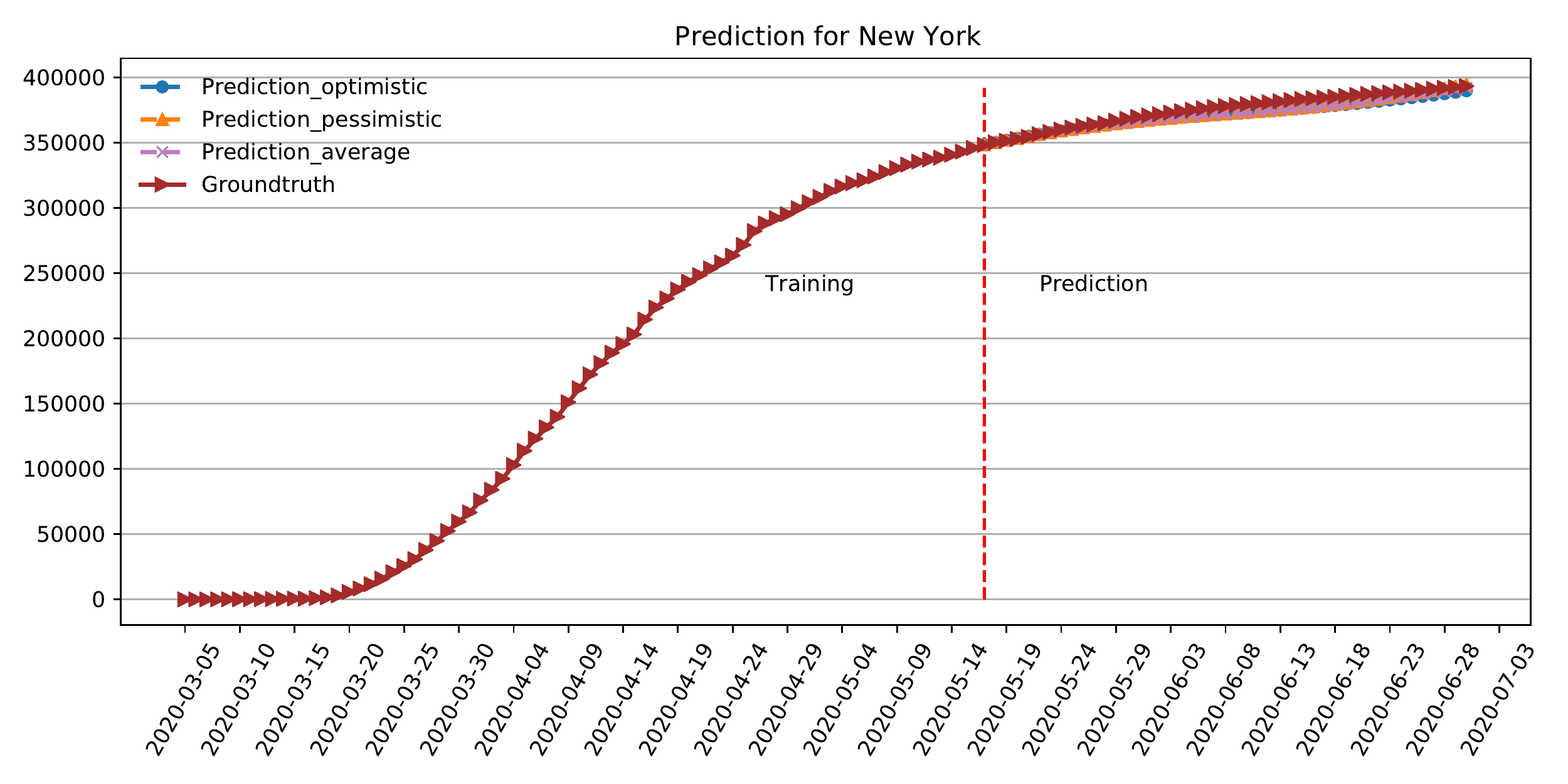}  
  \caption{New York. M = $0.22$.}
  \label{fig:sub-second3}
\end{subfigure}
\begin{subfigure}{.45\textwidth}
  \centering
  % include second image
  \includegraphics[width=1.\linewidth]{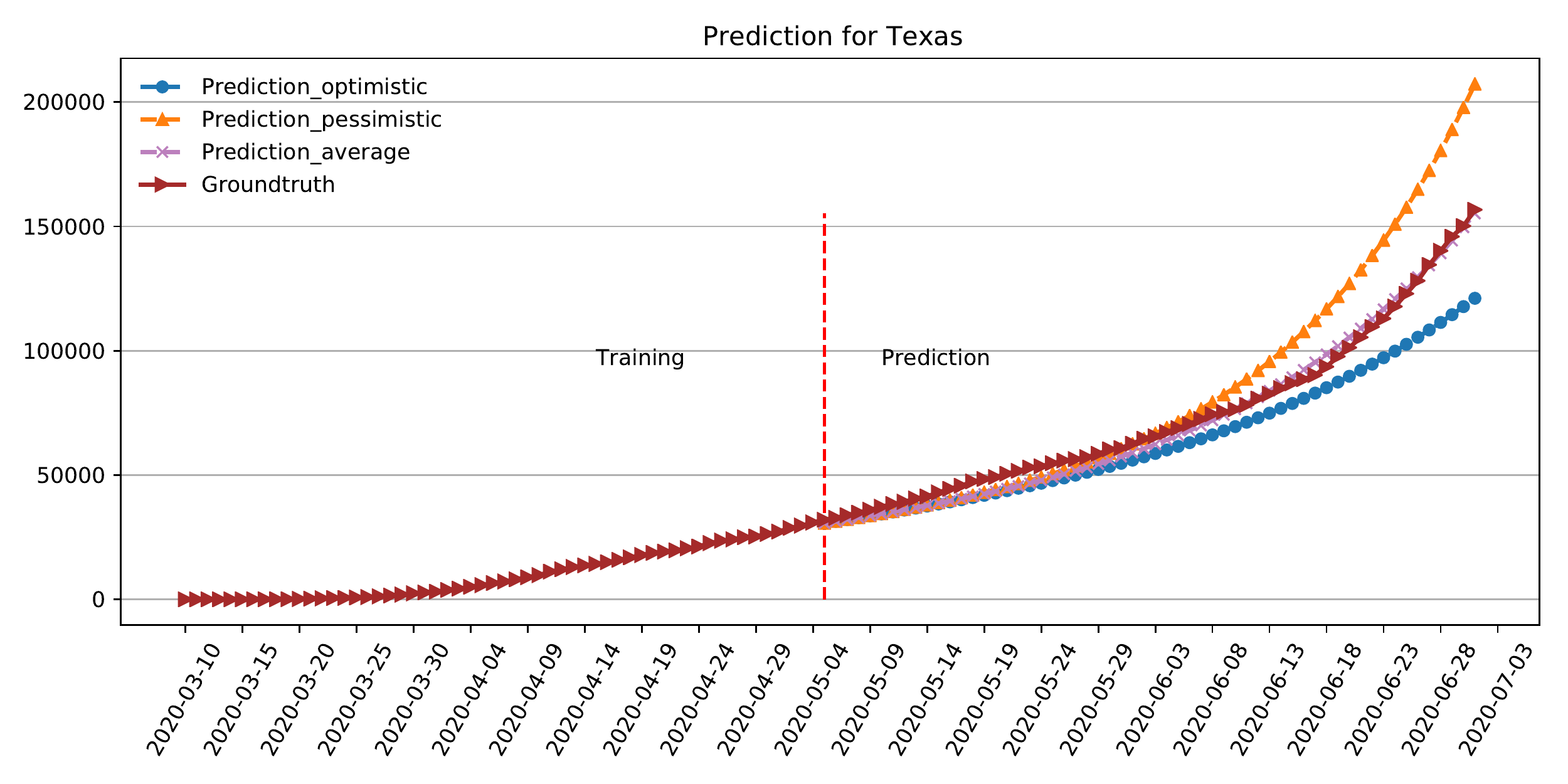}  
  \caption{Texas. M = $0.88$.}
  \label{fig:sub-second4}
\end{subfigure}

\caption{Model validation during reopening period.}
\label{fig:val}
\end{figure*}

\section{Model Validation Experiments} \label{sec:experiment}

For validation, we evaluate our model on four geographically and demographically diverse states in the US: California, Florida, New York and Texas. Readers may also refer to our public website\footnote{https://covid19predictions.csl.illinois.edu/} for up-to-date predictions for these and other states. 

To train our model, we need two calibration policies for each state, as described earlier. The first one we use is the ``no distancing'' policy that reflects the conditions before distancing was implemented. We start monitoring when the number of confirmed cases exceeds 10 and end when the first distancing policy is implemented. The second policy used for calibration is the initial ``strict distancing'' policy. It starts when distancing is first implemented and ends when re-opening starts. We then predict the effects of the reopening policy. Figure~\ref{fig:val} shows the prediction results, compared to ground truth. The pessimistic/optimistic prediction curves shown in the figure are generated by changing the predicted $\tau_{eq}$ up/down by $2\%$.

For each state, we show the date when the re-opening policy was implemented. Note that, we train the model using data collected before that date only (the training period). We then use the methodology presented earlier to compute a new model for the re-opening policy. We use that model to predict the impact of re-opening. Table~\ref{tb:reopening} summarizes the re-opening policies for the four states. Prediction results are shown through the end of June. We do not show July because many states changed their reopening policies again in July. For clarity. we restrict the predictions to the duration of one policy only.

The results shown in Figure~\ref{fig:val} indicate that our model provides a realistic and accurate prediction. The predicted trend can differ from the one during model training. For example, in Florida, one may notice that the cumulative cases before reopening (i.e., in the training period) have started to saturate. However, the prediction model for the re-opening phases correctly estimates that the number of cases will escalate again. Ground truth shown for that period confirms the prediction.

\begin{table}[h]
\caption{Reopening policy for each state.}
\label{tb:reopening}
\centering
\begin{tabular}{l|ccc}
\hline
 States & Reoping Policy \\ \hline
California  & Work: reopen 25\%, Commercial: reopen 50\% \\
%Georgia &  Work: reopen 50\%, Commercial: reopen 75\% \\
Florida & Work: reopen 50\%, Commercial: reopen 50\% \\
New York & Work: reopen 50\%, Commercial: reopen 50\% \\
Texas & Work: reopen 25\%, Commercial: reopen 75\% \\
\hline

\end{tabular}
\end{table}

Table~\ref{tb:R0} shows the value of the basic reproductive number for the four states under different policies. When the virus started, the observed $R_0$ for all the states was larger than $1$. Then, as a result of social distancing, the $R_0$ decreased significantly, especially for New York. As the states moved to the reopening period, the $R_0$ increased again. The values of California, Florida, and Texas exceed the safe threshold, which is consistent with the situation that they have experienced a second wave. 

\begin{table}[h]
\caption{Basic reproductive number $R_0$ under different policies.}
\label{tb:R0}
\centering
\begin{tabular}{l|ccc}
\hline
Policy                                       & No distancing &  Social distancing & Reopening  \\ \hline
California              & 1.83  & 1.07 & 1.16\\
Florida              & 2.77  & 0.67 & 1.03\\
%Georgia              & 2.37  & 0.72 & 0.97\\
New York              & 3.60  & 0.60 & 0.90\\
Texas                          & 2.50  & 0.85 & 1.06 \\ \hline
\end{tabular}
\end{table}

\begin{figure*}[ht]
\centering
\begin{subfigure}{.45\textwidth}
  \centering
  % include second image
  \includegraphics[width=1.\linewidth]{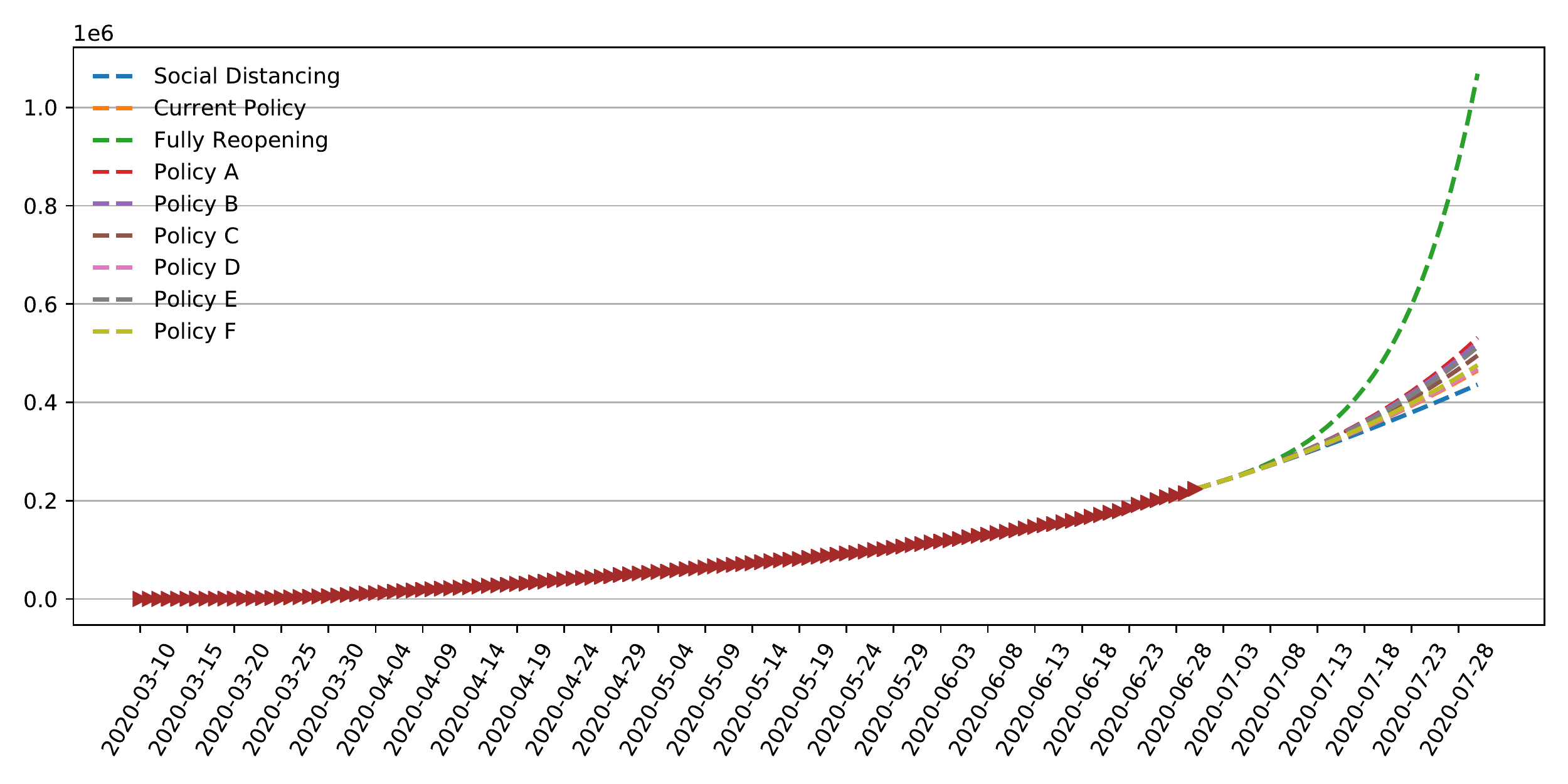}  
  \caption{California.}
  \label{fig:cawhatif}
\end{subfigure}
\begin{subfigure}{.45\textwidth}
  \centering
  % include first image
  \includegraphics[width=1.\linewidth]{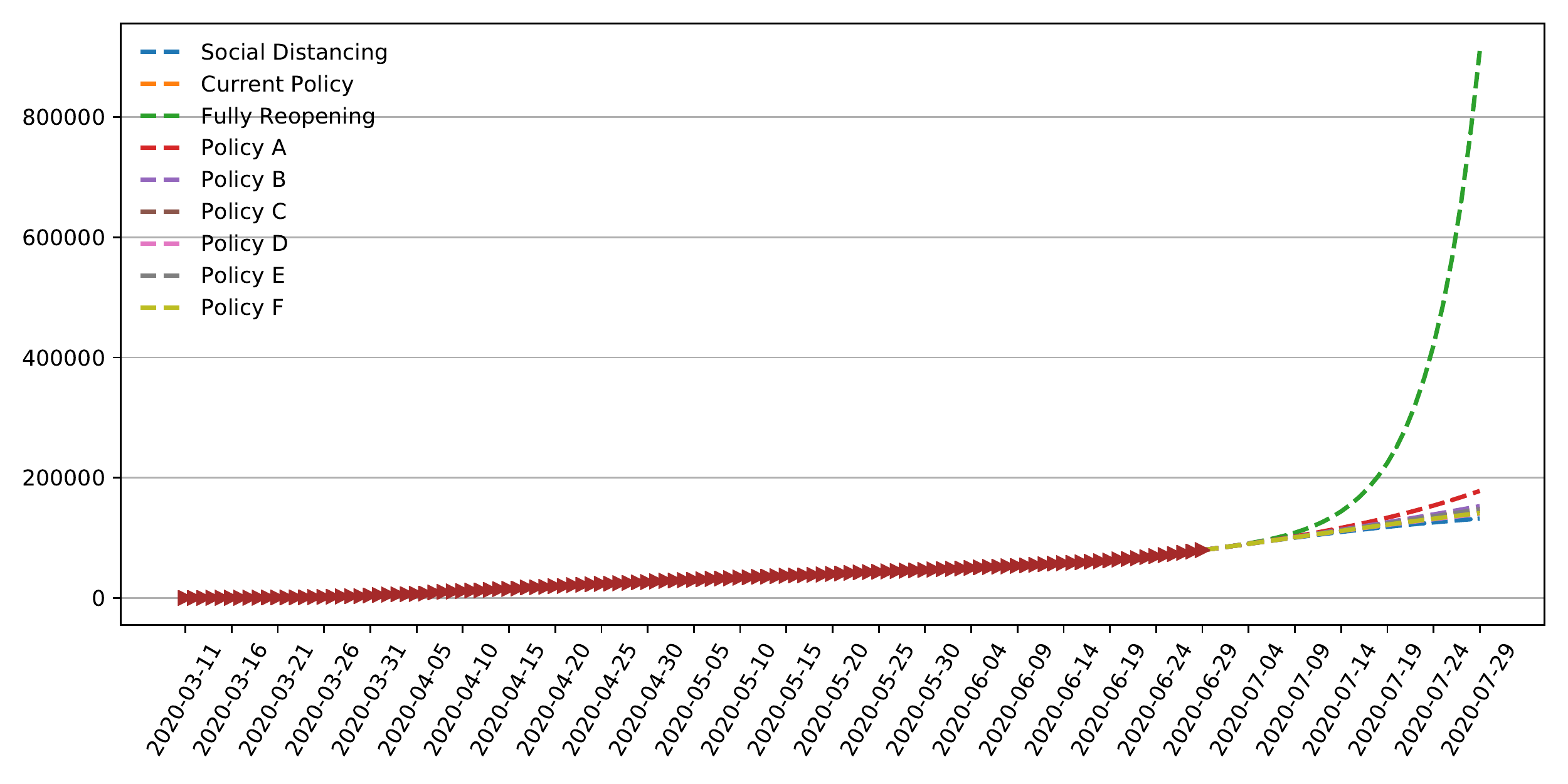}  
  \caption{Georgia.}
  \label{fig:geowhatif}
\end{subfigure}
\caption{Prediction for one month under different venue-based distancing policies.}
\label{fig:whatif}
\end{figure*}

\begin{figure*}[ht!]
\centering
\begin{subfigure}{.45\textwidth}
  \centering
  % include first image
  \includegraphics[width=1.\linewidth]{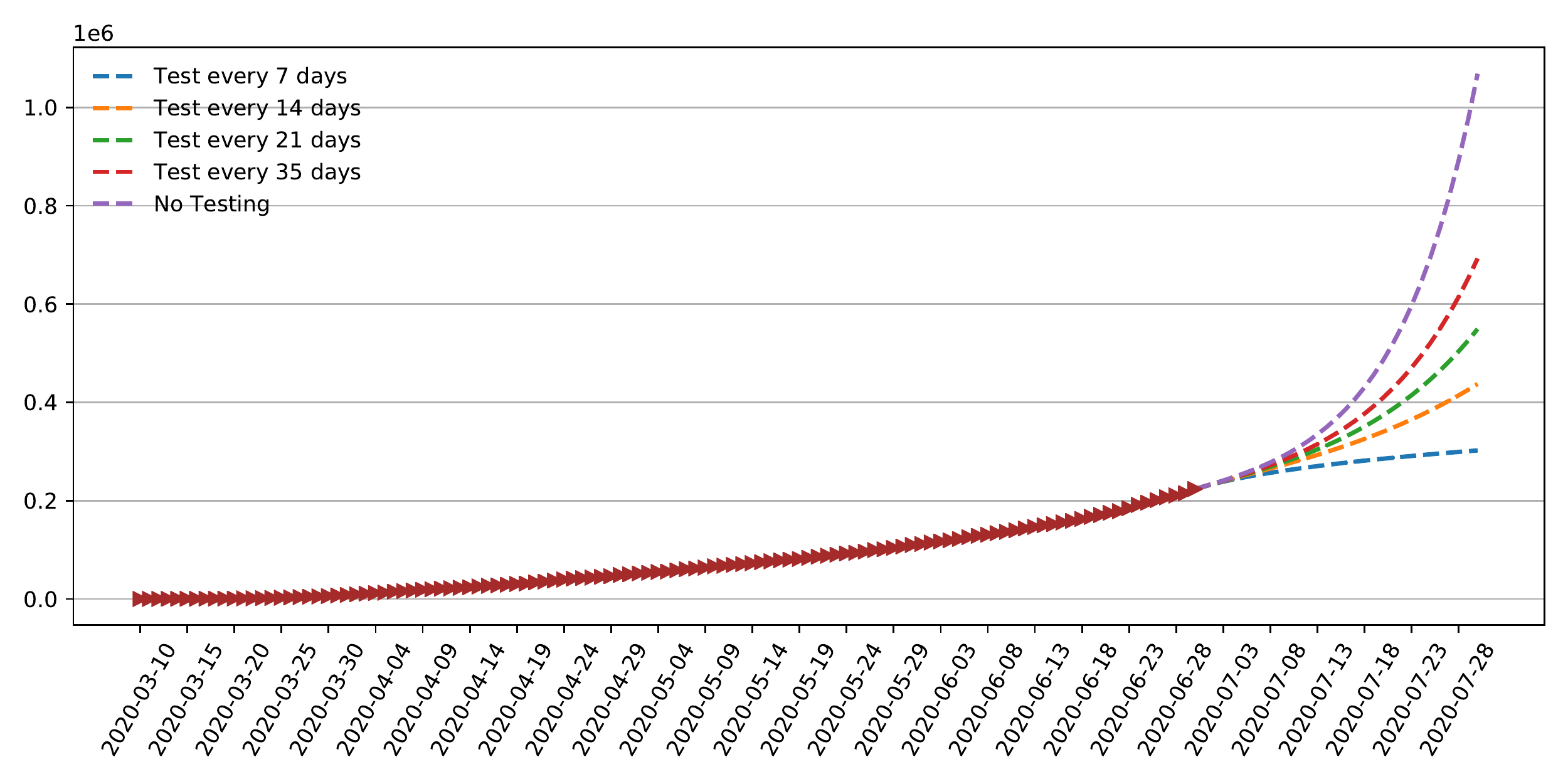}  
  \caption{California}
  \label{fig:sub-first}
\end{subfigure}
\begin{subfigure}{.45\textwidth}
  \centering
  % include second image
  \includegraphics[width=1.\linewidth]{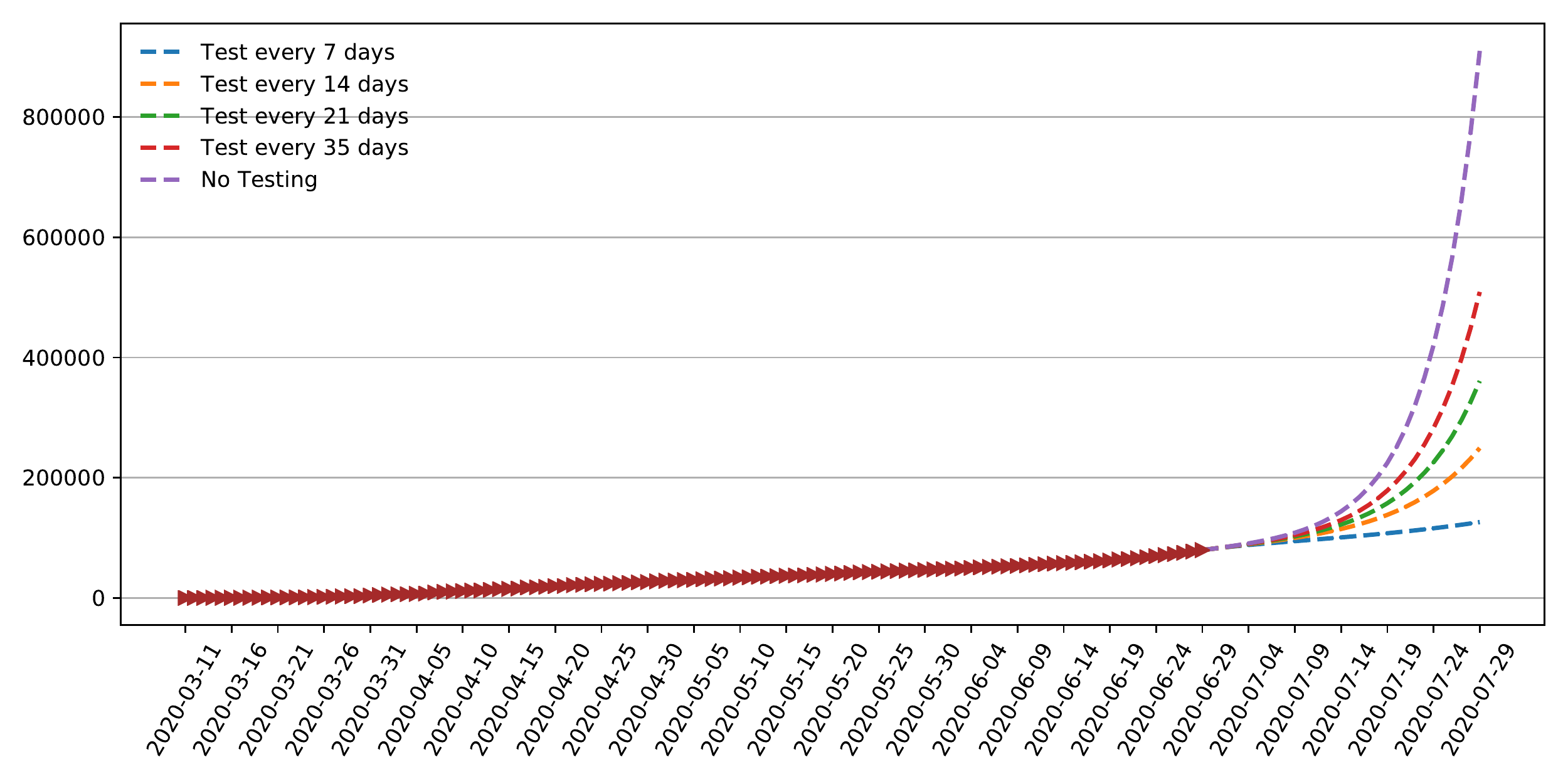}  
  \caption{Georgia}
  \label{fig:sub-second}
\end{subfigure}
\caption{Prediction for one month under different testing requirements in lieu of social distancing.}
\label{fig:testing}
\end{figure*}

\section{Re-opening Policies and Predicted Outcomes}

Next, we compare predicted efficacy of venue-based and testing-based policies for two example states, California and Georgia. We estimate the basic reproductive number $R_0$ and the number of cumulative cases after one month as two evaluation metrics. Readers can refer to  Section~\ref{sec:evaluation} for the used venue-based policies. The predicted curves for venue-based and testing-based policies are shown in Figure~\ref{fig:whatif} and Figure~\ref{fig:testing}, respectively. In testing-based policies, it is assumed that individuals are required to get tested periodically and that individuals found positive are quarantined. We vary the testing period from $7$ days to $35$. For the testing-based policies, we assume that no social distancing is used. To model this baseline, we modify a basic SIR model to feature a quarantined state. Let $L$ denote the required testing period. Thus, of those infected (I), the number of tested per day will be $(1/L) \cdot I$, on average. They will be quarantined. The resulting model becomes:

\begin{align}
    \frac{d S(t)}{dt} &= - \tau_{eq}\cdot f(t)\cdot S(t) I(t)
    \label{eq:SIR-S}\\
    \frac{d Q(t)}{dt} &= \frac{1}{L} \cdot I(t)
    \label{eq:SIR-Q}\\
    \frac{d I(t)}{dt} &= \tau_{eq}\cdot f(t)\cdot S(t) I(t) - \gamma \cdot I(t) - \frac{1}{L} \cdot I(t) 
    \label{eq:SIR-I}\\
    \frac{d R(t)}{dt} &= \gamma \cdot I(t),
    \label{eq:SIR-R}
\end{align}

The comparison shows that weekly testing offers better results than all compared venue-based policies. While it is difficult to build sufficient capacity for such frequent testing, this observation suggests that investments in cheap and widely available tests may be warranted.

% \begin{table}[h]
% \caption{Testing frequency that can complement a given reopening policy.}
% \label{tb:testing}
% \centering
% \begin{tabular}{c|c|c|c|c}
% \hline
% States   & \multicolumn{2}{c|}{California} & \multicolumn{2}{c}{Georgia} \\ \hline
% Policy   & L (days)     & $R_0$     & L (days)     & $R_0$   \\ \hline
% Policy A & 65      & 1.28                & 35    & 1.23             \\
% Policy B & 70      & 1.26                & 50    & 1.01              \\
% Policy C & 120      & 1.20                & 60   & 0.92              \\
% Policy D & 180    & 1.13               & 75    & 0.85              \\
% Policy E & 100      & 1.24                & 58    & 0.95              \\
% Policy F & 160      & 1.15                & 73    & 0.87              \\ \hline
% \end{tabular}
% \end{table}

%% file: 30-related.tex
\section{Related Work}
\label{sec:related}

The work is new in developing a {\em mesoscopic\/} epidemiological model in which the fundamental abstraction is neither the individual agent (node) nor an entire community, but rather an element in between: a single {\em broadcast channel\/} or {\em mixing domain\/}. Often popularity or size distribution of human and social artifacts follows standard profiles, such as Zipf Law~\cite{li2002zipf}, resulting in the same striking statistical distribution regularity in fields as diverse as linguistics~\cite{powers1998applications,dahui2005true}, urban populations~\cite{soo2005zipf,moura2006zipf}, business income distributions~\cite{okuyama1999zipf}, and the Internet~\cite{adamic2002zipf}. Recent executive orders in the US, in effect, manipulate the distribution of sizes of available social interaction venues by closing or reopening some mixing domains. Our mesoscopic model allows us to link social distancing policy decisions to the distribution of mixing domains, which in turn allows estimating future viral spread. The work aims to inform decision-making on crisis mitigation policies.   

Mesoscopic models are a middle ground between two well-populated extremes in current literature; namely, 
{\em agent-centric\/} models and {\em population-centric\/} models  (representing microscopic and macroscopic models, respectively). A recent survey discusses these existing models and the corresponding cascade mitigation policies they allow~\cite{nowzari2016analysis}. 
On one hand, agent-centric models~\cite{IC,realNetoworkEigenvalue,PhysRevE.63.066117,urbanNetwork} start with the behavior of individual agents, as well as their connectivity graphs. They enable reasoning about fine-grained mitigation strategies, such as inoculation of specific agents to reduce disease spread.  Agent-based models can also be used for detailed simulations to understand the impact of a large variety of detailed interventions. While very powerful and versatile, they require inputs that are difficult to collect, such as the interaction graph of all agents in the system. This limitation often renders them less suitable in practice. 

On the other hand, population-centric models, such as SIR~\cite{SIR}, SEIR~\cite{SEIR}, SIS~\cite{SIS}, and SQIS~\cite{SQIS}, focus on the total population. They reason about statistics of entire communities, such as the total number of infected, susceptible, and recovered individuals. They can also model high-level mitigation strategies such as inoculation of a specified fraction of the entire community (i.e., their removal from the susceptible list). These models, however, do not offer a clear way of reasoning about impacts of finer grained decisions, such as closure of some fraction of businesses or meeting venues. Macroscopic methods for COVID-19 trend prediction~\cite{NBERw26901,EARLY,TimeSIR,dataSIR,MLSIR,SEIR-covid19,Nesteruk2020.02.12.20021931} thus lack the ability to forecast effects of different policies.

To enable a more detailed analysis of epidemics, researchers extended the population-centric models by dividing the whole population into several groups, to form a finer-grained basis for analysis~\cite{complexNetworkEpidemic}. Heterogeneous mixing approaches~\cite{HeterogeneousEpidemic,YANG2007189,complexHeterogeneousEpidemic,Bogua2003} split individuals by their contact degrees. For a given degree, they use three differential equations to describe the evolution of three states. Age-structured epidemiological models \cite{singh20,ageSIR,Victor2020.03.28.20046300} assume different properties of people in different age groups. The epidemics are then governed by several sets of differential equations and some mixing policies. Some extensions~\cite{individualSIR} consider the state of each individual and study the state evolution during the infection process.

Unlike the above solutions that are based on partitioning of {\em individuals\/} or {\em communities\/} (by some demographic, geographic, of interaction-based pattern), we borrow inspiration from social media to focus on social mixing domains instead. Interactions need venues to facilitate them. The interaction patterns are thus a function of social mixing domains (i.e., the venues) that remain available.  
The advantage of our approach lies in its venue-centric model, as opposed to the more common community-centric or agent-centric models.

%% file: 60-conclusion.tex
\section{Conclusion}
\label{sec:conclusion}
The paper introduced a methodology for evaluating COVID-19 mitigation policies that reply on manipulation of available social {\em mixing domains\/}. The model relates virus transmissibility to the availability of social interaction venues, manipulated by social distancing policies. The results show that the simplified model is capable of accurately predicting changes in the contagion time series, allowing decision-makers to experiment with various distancing policies. For comparison, results of testing-based policies were included, showing that no distancing policies are as effective as regular testing.  

%% file: 40-model.tex
\section*{Appendix A}
Let the number of susceptible individuals in a community be denote by $S(t)$ and the number of infected individuals by $I(t)$. Let $N$ denote the total population, and $\beta$ the effective rate of spread.
Now assume that the region consists of $n$ social interaction venues. An individual divides their time among several venues (e.g., home, office, and other outlets). We call individuals who visit venue, $D_i$, {\em patrons of\/} $D_i$. Let $N_i = |D_i|$ be the nominal occupancy of $D_i$, which we call its size. Let $\eta_i$ be the average time a patron spends in $D_i$. 

Let us further denote the rate of transmission from {\em one\/} infected individual (to all susceptible individuals) per unit time (say one day) in venue $D_i$ by $\beta_i$. Let the expected number of susceptible and infected individuals in the $i$-th venue, at time $t$, be denoted by $S_i(t)$ and $I_i(t)$, respectively. Furthermore, let us define $\zeta_i(t) = S_i(t)/S(t)$ and $\kappa_i(t) = I_i(t)/I(t)$. The fractions, $\zeta_i(t)$ and $\kappa_i(t)$ are, respectively, the expected fraction of all those susceptible and the expected fraction of all those infected, who are patrons of venue $D_i$. Assuming an SIR model,
we can thus write:

\begin{align}
    \frac{d S_i(t)}{dt} &= - \beta_i \eta_i\cdot S_i(t)I_i(t)~ /~ N_i, 
    \label{eq:S1}\\
    \frac{d I_i(t)}{dt} &= \beta_i \eta_i\cdot S_i(t)I_i(t) ~/~ N_i - \gamma \cdot I_i(t),
    \label{eq:S2}\\
    \frac{d R_i(t)}{dt} &= \gamma \cdot I_i(t),
    \label{eq:S3}
\end{align}

\noindent
The above equations roughly assume that each infected individual in $D_i$ makes, say, $c_i$ encounters in the venue per unit time, of which, therefore, $c_i S_i(t)/N_i$ are in susceptible population ($S_i(t)/N_i$ is the probability of the susceptible in $D_i$) . If the probability of transmission per encounter is $\tau_i$, then each infected individual passes the virus to $\tau_i c_i S_i(t)/N_i$ others, leading to the above equations, where $\beta_i = \tau_i c_i$. In our analysis, we assume that the number of other encountered individuals in a grows with the size of the venue (e.g., one passes more people in a conference than in a small party). Thus, $c_i$ grows proportionally to $N_i$, whereas $\tau_i$ (which can be redefined to absorb the proportionality constant) is generally higher for smaller venues, since people tend to have closer (and/or longer) encounters in smaller groups. Thus, we can rewrite $\beta_i$ as:

\begin{equation}
    \beta_i = \tau_i N_i
\end{equation}

where $\tau_i$ is transmissibility (per encounter with another individual) within a venue, which tends to be higher (due to closer and longer encounters) for smaller venues. Substituting in Equations~(\ref{eq:S1}), (\ref{eq:S2}), and (\ref{eq:S3}), we thus get:

\begin{align}
    \frac{d S_i(t)}{dt} &= - \tau_i \eta_i\cdot S_i(t)I_i(t), \\
    \frac{d I_i(t)}{dt} &= \tau_i \eta_i\cdot S_i(t)I_i(t) - \gamma \cdot I_i(t),\\
    \frac{d R_i(t)}{dt} &= \gamma \cdot I_i(t),
\end{align}

Adding up over all venues, we get:

\begin{align}
    \frac{d S(t)}{dt} &= - \sum_{i=1}^n\tau_i \eta_i\cdot S_i(t)I_i(t) = - \left( \sum_{i=1}^n \tau_i \eta_i \zeta_i \kappa_i \right) S(t) I(t)
    \label{eq:SIR1}\\
    \frac{d I(t)}{dt} &= \left( \sum_{i=1}^n \tau_i \eta_i \zeta_i \kappa_i \right) S(t) I(t) - \gamma \cdot I(t) 
    \label{eq:SIR2}\\
    \frac{d R(t)}{dt} &= \gamma \cdot I(t),
    \label{eq:SIR3}
\end{align}

\paragraph{Perfect Mixing Approximation.}Let us briefly discuss the implications of the above equations. In a system, where everyone is restricted to venues that are perfectly quarantined (for example, restricted to their family residences under strict quarantine), the ratio of infected in the quarantine zone, $\kappa_i = I_i(t)/I(t)$, will be disproportionately higher than what venue size might predict this ratio to be (i.e., $I_i(t)/I(t) > N_i/N$). Similarly, the same ratio outside the quarantine zone will be lower. In our analysis, however, we assume that strict quarantine is no longer socially viable. Instead, individuals from different venues will mix (in other venues). For example, individuals from different households might mix in the same grocery store or office and individuals from different offices might mix at the same bus stop. (Of course, the opportunities for mixing are constrained by available venues.) We assume that mixing fails to localize infections in any subset of venues, and instead spreads the infection as broadly as possible. The above mixing assumption leads to an important {\em worst-case approximation\/}. Namely, if mixing is perfect, the expected number of susceptible (infected) individuals in a venues is roughly proportional to its size. More specifically:

\begin{align}
S_i (t) & \approx  \frac {N_i}{N} S(t)\\
I_i (t) & \approx  \frac {N_i}{N} I(t)
\end{align}

We consider this a worst-case approximation because the resulting analysis tends to maximize estimates of total spread. If mixing is imperfect, then virus spread will slow down sooner in more heavily impacted venues (due to scarcity of remaining susceptible individuals), while it will also proceed at a slower rate in others (due to scarcity of infected individuals). The worst-case assumption is helpful from the perspective of erring on the safe side. Let us define $\alpha_i$ to be the fraction $\alpha_i = N_i/N$. From the above, we get:

\begin{equation}
    \zeta_i(t) = \kappa_i (t) = \alpha_i
    \label{eq:approx}
\end{equation}

Substituting from Equation~(\ref{eq:approx}) into Equations~(\ref{eq:SIR1}), (\ref{eq:SIR2}), and (\ref{eq:SIR3}), we get:
\begin{align}
    \frac{d S(t)}{dt} &= - \tau_{eq} S(t) I(t)
    \label{eq:SIR-a}\\
    \frac{d I(t)}{dt} &= \tau_{eq} S(t) I(t) - \gamma \cdot I(t) 
    \label{eq:SIR-b}\\
    \frac{d R(t)}{dt} &= \gamma \cdot I(t),
    \label{eq:SIR-c}
\end{align}

where:

\begin{equation}
    \tau_{eq} =  \sum_{i=1}^n \tau_i \eta_i \alpha_i^2 
    \label{eq:beta}
\end{equation}

Observe that the above equations have the same form as the basic SIR model. Thus, $\tau_{eq}$ can be interpreted as the equivalent transmissibility of the virus by considering all venues.
The above equations, in fact, are a proper generalization of an SIR model to the case of multiple interaction venues. 

Note that, Equation~(\ref{eq:beta}) has the interesting property that infection transmissibility, $\tau_{eq}$, depends on the overall {\em distribution\/} of quantities $\tau_i$, $\eta_i$, and $\alpha_i$, across the venues. By manipulating some of the venues (e.g., closing them), one can thus reduce the equivalent transmissibility, $\tau_{eq}$, of the disease.

Now let us break the local population into a set $\mathcal{G}$ of non-overlapping groups. Let $\eta_{ij}$ denote the average amount of time that 
individuals in group $g_j \in \mathcal{G}$ spend in venues $i$ (some fractions could be zero). Also, let $f_{ij}$ be the average fraction of occupancy of venues $i$ who are from group $g_j$, and let $\tau_{ij}$ be the probability that an infected person in venues $i$ who encounters a member of group $g_j$ successfully infects them (i.e., transmissibility). 
One can thus approximately rewrite the product $\tau_i \eta_i$ as a weighted sum of per-group products:

\begin{equation}
    \tau_i \eta_i = \sum_{j \in \mathcal{G}}  (\tau_{ij} \eta_{ij}) f_{ij}
    \label{eq:weighted}
\end{equation}

Substituting from Equation~(\ref{eq:weighted}) into Equation~(\ref{eq:beta}), we get: 

\begin{equation}
    \tau_{eq} =  \sum_{i=1}^n \sum_{j \in \mathcal{G}} f_{ij} ~ \tau_{ij} ~ \eta_{ij} ~ \alpha_i^2 
    \nonumber
    % \label{eq:theorem}    
\end{equation}

This equation is the statement of the mixing theorem \qedsymbol

\subsection*{Appendix B: The Mitigation Policy}
\label{sec:mitigation}

Consider some mitigation policy, $\mathcal{P}$. Let $\eta'_i$ and $\alpha'_i$ be the fractional amount of time and fractional membership of the domain after the policy is implemented. Thus:

\begin{equation}
    \tau'_{eq} =  \sum_{i=1}^n \tau'_i \eta'_i {\alpha'_i}^2 
    \label{eq:beta2}
\end{equation}

In general, announced social distancing measures might not come into effect immediately. In the model, we may consider an exponential function $f(t)$ by which $\tau_{eq}$ changes to $\tau'_{eq}$. Thus, when a new policy is announced, starting at time $t_0$, we can rewrite:

\begin{align}
    \frac{d S(t)}{dt} &= -\tau_{eq}\cdot f(t) \cdot S(t)I(t), \\
    \frac{d I(t)}{dt} &= \tau_{eq}\cdot f(t) \cdot S(t)I(t) - \gamma \cdot I(t),\\
    \frac{d R(t)}{dt} &= \gamma \cdot I(t), 
\end{align}

where:

\begin{equation}
    f(t) =\left\{  
             \begin{array}{rcl}
             1, && t < t_0 \\
             \frac{\tau'_{eq}}{\tau_{eq}} + (1-\frac{\tau'_{eq}}{\tau_{eq}})\cdot e^{-c_1 (t-t_0)}, && t \geq t_0 \\
             \end{array}  
    \right.
\end{equation}

In our model, the parameters $\tau_{eq}$ and $\gamma$ can be estimated from past time-series data of the contagion cascade (say, before a mitigation policy is implemented). After the first change in policy, parameter $c_1$ can generally be assumed, as it is the inverse of the convergence time-constant. The fraction $\frac{\tau'_{eq}}{\tau_{eq}}$ is then computed from Equations~(\ref{eq:beta}) and (\ref{eq:beta2}), by accounting for the implemented policy. To do so, it is useful to remember that $\eta_i$ and $\eta'_i$ represent the average fraction of time an individual spends in venue $D_i$, before and after a mitigation policy is implemented.